\documentclass[aps,prl,reprint,superscriptaddress]{revtex4-2}

\usepackage{amsmath,amssymb,bm}
\usepackage{physics}
\usepackage{graphicx}

\begin{document}

\title{Information-Geometric bound on the robustness of entanglement generation}

\author{Zain H. Saleem}
\affiliation{Mathematics and Computer Science Division, Argonne National Laboratory, Lemont, Illinois 60439, USA}

\begin{abstract}
Entanglement generation is a central resource for quantum information processing, quantum networking, and quantum sensing. In practical implementations, however, entangling interactions are inevitably subject to uncertainty and fluctuations in the interaction strength. We investigate the robustness of entanglement generation in the presence of such imperfections and establish a direct connection between the robustness of entanglement generation and quantum Fisher information (QFI). For two interacting qubits, we show that the reduction in concurrence caused by fluctuations in the interaction parameter is bounded by the QFI with respect to the interaction strength.
\end{abstract}

\maketitle

\textit{Introduction:} 
Entanglement is a fundamental resource in quantum information science \cite{NielsenChuang2000,Ladd2010,Kimble2008,Bernien2013,Humphreys2018,Giovannetti2004,Pezze2018}, and a variety of measures have been developed to quantify it, including entanglement entropy, entanglement of formation, negativity, and robustness-based measures \cite{Horodecki2009,VidalTarrach1999}. For two-qubit systems, concurrence provides a particularly useful measure due to its analytical tractability and its direct connection to entanglement of formation \cite{HillWootters1997,Wootters1998}. Another quantity of central importance is the quantum Fisher information (QFI), which quantifies the sensitivity of a quantum state to changes in a parameter and determines the ultimate precision achievable in parameter estimation through the quantum Cram'er--Rao bound \cite{Helstrom1976,Holevo1982,BraunsteinCaves1994,Paris2009}. Beyond its metrological significance, the QFI also admits an information-geometric interpretation through its connection to the Bures distance \cite{BraunsteinCaves1994,Paris2009}. Recently, it was shown that for two-qubit systems evolving under entangling interactions, the parametric curvature of concurrence is bounded by the QFI associated with the same interaction parameter \cite{SaleemCoE2025}. This connection provides a natural starting point for investigating the robustness of entanglement generation in the presence of fluctuations in the entangling interaction.

Fluctuations in the parameters that control two-qubit entanglement are not merely a theoretical concern but exists in all experimental quantum platforms. In trapped-ion two-qubit gates, the entangling operation is mediated by collective motion and driven by Raman beams; experimental error budgets explicitly include motional heating and dephasing, spin dephasing, and Raman-beam intensity drift, with measured Rabi-frequency fluctuations along the two-qubit gate beam paths \cite{Ballance2016}. Related high-fidelity ion-gate experiments emphasize improved laser-beam quality and control as part of suppressing two-qubit gate infidelity \cite{Gaebler2016}. In neutral-atom Rydberg platforms, two-qubit controlled-phase gates are implemented by shaping the Rabi frequency $\Omega(t)$ and phase $\phi(t)$ of the driving laser, so fluctuations in laser amplitude, phase, detuning, or atomic motion directly perturb the entangling operation \cite{Levine2018,Evered2023}. In two-qubit quantum sensing and noise spectroscopy, experiments with superconducting qubits have explicitly reconstructed spatiotemporally correlated stochastic noise using two qubits coupled to a shared engineered noise source, illustrating the importance of fluctuating two-qubit environments for both computation and metrology \cite{vonLupke2020}. Similar stochastic effects arise in quantum networking experiments. In the landmark demonstration of heralded entanglement between distant solid-state spin qubits, entanglement generation relied on spin--photon entanglement and photon interference, making the resulting two-qubit state sensitive to optical phase stability and photon indistinguishability \cite{Bernien2013}. More recently, Hermans \emph{et al.} performed a detailed theoretical and experimental analysis of remote two-qubit entanglement generation and showed that optical phase differences, frequency offsets between remote qubits, imperfect excitation, and other experimental imperfections directly affect the phase and fidelity of the generated entangled state \cite{Hermans2023}. In these settings, fluctuations do not necessarily appear as variations of a microscopic coupling constant, but rather as fluctuations in the effective parameter governing entanglement generation. These observations naturally motivate the question of how robust entanglement generation is to uncertainty in the parameter controlling the entangling interaction.

Motivated by these considerations, we investigate the robustness of entanglement generation in the presence of fluctuations in the interaction strength. We consider the simplest nontrivial setting of two interacting qubits and ask how uncertainty in the entangling interaction affects the amount of entanglement that can be generated. To quantify this effect, we introduce a measure of robustness based on the reduction in concurrence induced by interaction-strength fluctuations. We show that, in the weak-fluctuation regime, this quantity is directly related to the local curvature of concurrence with respect to the interaction parameter. Combining this observation with recently established connections between concurrence curvature and quantum Fisher information \cite{SaleemCoE2025}, we derive an information-geometric bound that links the robustness of entanglement generation to the quantum Fisher information associated with estimating the interaction strength. Our results provide a new operational interpretation of quantum Fisher information and establish a direct connection between entanglement generation, imperfect control, and quantum metrology.

\textit{Canonical two-qubit interaction.}
We consider two interacting qubits governed by a Hamiltonian of the form
\begin{equation}
H(g)=gh,
\label{eq:H}
\end{equation}
where \(g\) denotes the interaction strength and \(h\) is a dimensionless interaction operator. Throughout this work we focus on pure-state dynamics generated by two-qubit interactions and investigate the relationship between entanglement generation and metrological sensitivity with respect to the same interaction parameter. Two-qubit systems constitute the simplest setting in which this connection can be explored. While single-qubit Hamiltonians can generate coherence and parameter sensitivity, they cannot generate entanglement. In contrast, two-qubit interactions simultaneously generate quantum correlations and encode information about the interaction strength into the evolving quantum state. The interaction parameter therefore plays a dual role: it governs both the amount of entanglement produced and the distinguishability of neighboring quantum states.

The most general two-qubit interaction Hamiltonian can be written as
\begin{equation}
h=\sum_{j,k=x,y,z}\eta_{jk}\sigma_j\otimes\sigma_k,
\label{eq:generalh}
\end{equation}
where \(\eta_{jk}\) is a real coefficient matrix. Since concurrence and quantum Fisher information are invariant under local unitary transformations, the comparison between these quantities depends only on the nonlocal content of the interaction. Consequently, the coefficient matrix \(\eta\) may be diagonalized through a singular-value decomposition and the associated local rotations absorbed into a redefinition of the computational basis. Any two-qubit interaction can therefore be reduced to the canonical form
\begin{equation}
h=
\eta_x\sigma_x\otimes\sigma_x+
\eta_y\sigma_y\otimes\sigma_y+
\eta_z\sigma_z\otimes\sigma_z,
\label{eq:canonicalh}
\end{equation}
without affecting either concurrence or QFI. This reduction was used extensively in Ref.~\cite{SaleemCoE2025} and allows both quantities to be expressed analytically while retaining complete generality.

The Bell states $|\beta_{ab}\rangle
=
\frac{|0,b\rangle+(-1)^a|1,\bar b\rangle}{\sqrt2},
\qquad
a,b\in\{0,1\},
$
form an eigenbasis of Eq.~(\ref{eq:canonicalh}). The corresponding eigenfrequencies are $\omega_{ab}
=
(-1)^a\eta_x
-
(-1)^{a+b}\eta_y
+
(-1)^b\eta_z.
\label{eq:omegas}
$
An arbitrary pure initial state may therefore be expanded as
\begin{equation}
|\Psi_0\rangle
=
\sum_{a,b=0,1}
\beta_{ab}
|\beta_{ab}\rangle,
\label{eq:bell_expansion}
\end{equation}
and evolves according to
\begin{equation}
|\Psi(g,t)\rangle
=
\sum_{a,b}
\beta_{ab}
e^{-ig\omega_{ab}t}
|\beta_{ab}\rangle.
\label{eq:evolved_state}
\end{equation}

The interaction dynamics are completely determined by two ingredients: the Bell-state amplitudes \(\beta_{ab}\) and the interaction frequencies \(\omega_{ab}\). The former determine how strongly each Bell sector participates in the evolution, while the latter determine the rates at which relative phases accumulate. This decomposition provides a particularly transparent description of both entanglement generation and parameter estimation because all quantities of interest can be expressed directly in terms of the same set of interaction frequencies.

\vspace{1ex}

\textit{Concurrence}
To quantify the entanglement generated by the interaction we employ concurrence, which admits a particularly simple analytical representation for two qubits and is therefore ideally suited for investigating how entanglement depends on the interaction strength. In Ref.~\cite{SaleemCoE2025} the analytical form of concurrence is derived for the state given by equation (\ref{eq:evolved_state}),
\begin{equation}
C(g,t)
=
\left|
\sum_{a,b}
(-1)^{a+b}
\beta_{ab}^{\,2}
e^{-2ig\omega_{ab}t}
\right|.
\label{eq:Cgeneral}
\end{equation}
Equation~(\ref{eq:Cgeneral}) plays a central role in everything that follows. It expresses concurrence as a coherent superposition of interaction-frequency sectors. The generated entanglement is therefore not determined solely by the Bell-state populations \(|\beta_{ab}|^2\), but also by the relative phases accumulated between different frequency components during the evolution.

This observation provides an intuitive picture of entanglement generation. Each Bell component evolves at a frequency \(\omega_{ab}\), producing relative phases proportional to \(g\omega_{ab}t\). Concurrence measures how these phase factors interfere. Constructive interference between the different sectors enhances entanglement generation, whereas destructive interference suppresses it. Entanglement generation may therefore be viewed as a frequency-interference phenomenon whose structure is entirely encoded in the interaction spectrum.

The representation (\ref{eq:Cgeneral}) is especially useful because it makes explicit how the interaction parameter enters the concurrence. Any uncertainty in the interaction strength modifies the phases appearing in Eq.~(\ref{eq:Cgeneral}), thereby changing the interference pattern responsible for entanglement generation. The robustness of entanglement generation can therefore be understood in terms of the sensitivity of these interference terms to perturbations of the interaction parameter.

\vspace{1ex}

\textit{Quantum Fisher information}
The sensitivity of the quantum state itself to variations in the interaction strength is quantified by the quantum Fisher information. For a pure state evolving under Eq.~(\ref{eq:H}), the QFI with respect to \(g\) is
\begin{equation}
F_Q^{(g)}
=
4t^2(\Delta h)^2,
\label{eq:qfi_def}
\end{equation}
where
\begin{equation}
(\Delta h)^2
=
\langle h^2\rangle
-
\langle h\rangle^2
\end{equation}
is evaluated in the initial state. The QFI determines the ultimate precision with which the interaction strength may be estimated through the quantum Cram\'er--Rao bound and provides a local measure of the distinguishability of neighboring quantum states.

Using the Bell-state expansion of Eq.~(\ref{eq:bell_expansion}), the QFI assumes the explicit form
\begin{equation}
F_Q^{(g)}
=
4t^2
\left[
\sum_{a,b}
|\beta_{ab}|^2\omega_{ab}^2
-
\left(
\sum_{a,b}
|\beta_{ab}|^2\omega_{ab}
\right)^2
\right].
\label{eq:qfigeneral}
\end{equation}
Unlike concurrence, Eq.~(\ref{eq:qfigeneral}) depends only on the variance of the interaction spectrum sampled by the initial state. Relative phases do not appear. The QFI therefore quantifies the spread of interaction frequencies rather than their interference.

This distinction makes the existence of a direct relationship between concurrence and QFI highly nontrivial. Concurrence depends on coherent phase interference between Bell sectors, whereas the QFI depends only on the spectral variance of those same sectors. One quantity characterizes entanglement generation, while the other characterizes parameter sensitivity. Nevertheless, both originate from the same interaction spectrum, suggesting that a deeper connection may exist.

From an information-theoretic perspective, large values of the QFI imply that a small change in interaction strength produces a large displacement in Hilbert space. Since the same interaction is also responsible for generating entanglement, it is natural to ask whether the sensitivity of the quantum state constrains the sensitivity of the entanglement generated by that state. This question motivates the introduction of the curvature of entanglement.

\vspace{1ex}

\textit{Curvature of entanglement and information-geometric bound.}
To place concurrence and QFI on equal footing, Ref.~\cite{SaleemCoE2025} introduced the curvature of entanglement (CoE),
\begin{equation}
\mathrm{CoE}(g,t)
=
-
\frac{\partial^2 C(g,t)}
{\partial g^2}.
\label{eq:coe}
\end{equation}
The CoE quantifies the local sensitivity of entanglement generation to variations in the interaction strength. Whereas concurrence measures the amount of entanglement generated at a given operating point, the CoE measures how rapidly that entanglement changes as the interaction parameter is varied.

The use of a second derivative is particularly natural in the present setting. At local maxima of concurrence the first derivative necessarily vanishes, rendering the second derivative the leading quantity governing the response of the generated entanglement to small parameter perturbations. The CoE therefore provides a local measure of the fragility of entanglement generation with respect to variations in the interaction strength. Large values correspond to sharply peaked entanglement landscapes, while small values correspond to comparatively flat landscapes.

Using the analytical representation (\ref{eq:Cgeneral}), it was shown in Ref.~\cite{SaleemCoE2025} that the curvature of entanglement satisfies the information-geometric bound
\begin{equation}
\mathrm{CoE}(g,t)
\le
F_Q^{(g)}(g,t),
\label{eq:coebound}
\end{equation}
for arbitrary pure two-qubit states evolving under Eq.~(\ref{eq:canonicalh}). The proof will not be reproduced here. Briefly, both quantities can be expressed in terms of the same centered interaction frequencies, with the QFI determined by the corresponding spectral variance and the CoE determined by a coherent combination of the same spectral fluctuations.

To state the saturation conditions, define the centered frequencies
$\Delta_{ab}=\omega_{ab}-\sum_{a,b}|\beta_{ab}|^2\omega_{ab}$
and
$X_{ab}=(-1)^{a+b}\beta_{ab}^{\,2}e^{-2igt\Delta_{ab}}$.
The first saturation condition requires constructive interference of all nonzero frequency sectors,
$\arg(\Delta_{ab}^{\,2}X_{ab})
=\arg(\Delta_{a'b'}^{\,2}X_{a'b'})$,
for all contributing sectors. Writing the concurrence amplitude in polar form as
$\widetilde C(g,t)=M(g,t)e^{i\phi(g,t)}$,
the remaining saturation conditions are
$\partial_g\phi=0$,
$\partial_g^2\phi=0$,
and
$\partial_g^2M\le0$. Under these conditions the information-geometric bound is saturated and one obtains $\mathrm{CoE}(g,t)
=
F_Q^{(g)}(g,t)$.

The results summarized above establish a direct relationship between entanglement generation and metrological sensitivity for deterministic interaction strengths. In realistic implementations, however, the interaction parameter is typically subject to stochastic fluctuations arising from calibration errors, control noise, and environmental perturbations. We now show that the same information-geometric structure underlying Eq.~(\ref{eq:coebound}) also governs the robustness of entanglement generation in the presence of interaction-strength uncertainty.

\textit{Robustness of entanglement generation.}
The results of the previous section characterize the response of concurrence to infinitesimal deterministic variations of the interaction strength. In realistic implementations, however, the interaction parameter is typically subject to stochastic fluctuations arising from calibration errors, control noise, and environmental perturbations. It is therefore natural to ask how robust the generated entanglement is to uncertainty in the interaction strength itself.

To address this question, we replace the intended interaction strength \(g\) by
\begin{equation}
g \rightarrow g+\xi,
\end{equation}
where \(\xi\) is a random variable drawn from a probability distribution \(p(\xi)\) satisfying
\begin{equation}
\int d\xi\, p(\xi)\,\xi =0,
\qquad
\int d\xi\, p(\xi)\,\xi^2=\sigma_g^2.
\end{equation}
For a given realization of the fluctuation, the generated concurrence is \(C(g+\xi,t)\). The experimentally observed concurrence is therefore the ensemble-averaged quantity
\begin{equation}
\overline C(g,t)
=
\int d\xi\, p(\xi)\, C(g+\xi,t).
\label{eq:Cbar}
\end{equation}

Motivated by the reduction in observed entanglement caused by such fluctuations, we define the robustness of entanglement generation with respect to interaction-strength uncertainty as
\begin{equation}
\mathcal R_g(g,t)
=
\frac{
C(g,t)-\overline C(g,t)
}
{\sigma_g^2}.
\label{eq:Rg}
\end{equation}
The quantity \(\mathcal R_g\) measures the reduction in concurrence normalized by the variance of the interaction-strength fluctuations. Small values of \(\mathcal R_g\) correspond to operating points that are relatively insensitive to uncertainty in \(g\), while large values indicate fragile operating points where even small fluctuations can significantly reduce the generated entanglement.

To relate Eq.~(\ref{eq:Rg}) to the curvature of entanglement introduced in the previous section, we consider the weak-fluctuation regime and expand the concurrence about the intended interaction strength,
\begin{equation}
C(g+\xi,t)
=
C(g,t)
+
\xi
\frac{\partial C}{\partial g}
+
\frac{\xi^2}{2}
\frac{\partial^2 C}{\partial g^2}
+
O(\xi^3).
\label{eq:Taylor}
\end{equation}
Substituting Eq.~(\ref{eq:Taylor}) into Eq.~(\ref{eq:Cbar}) and using the first two moments of the fluctuation distribution yields
\begin{equation}
\overline C(g,t)
=
C(g,t)
+
\frac{\sigma_g^2}{2}
\frac{\partial^2 C}{\partial g^2}
+
O(\sigma_g^3).
\label{eq:CbarTaylor}
\end{equation}
Using the definition of the curvature of entanglement,
\begin{equation}
\mathrm{CoE}(g,t)
=
-\frac{\partial^2 C(g,t)}{\partial g^2},
\end{equation}
we obtain
\begin{equation}
\lim_{\sigma_g\rightarrow0}
\mathcal R_g(g,t)
=
\frac{1}{2}
\mathrm{CoE}(g,t).
\label{eq:RgCoE}
\end{equation}
Equation~(\ref{eq:RgCoE}) provides an operational interpretation of the curvature of entanglement. The CoE was originally introduced as a geometric quantity characterizing the local curvature of the concurrence landscape. Equation~(\ref{eq:RgCoE}) shows that it also governs the leading-order degradation of experimentally observed entanglement caused by interaction-strength fluctuations. In this sense, the curvature of entanglement quantifies the weak-noise robustness of entanglement generation.

Combining Eq.~(\ref{eq:RgCoE}) with the information-geometric bound Eq.~(\ref{eq:coebound}), immediately yields
\begin{equation}
\lim_{\sigma_g\rightarrow0}
\mathcal R_g(g,t)
\le
\frac{1}{2}
F_Q^{(g)}(g,t).
\label{eq:mainbound}
\end{equation}
Equation~(\ref{eq:mainbound}) establishes that the fluctuation-normalized loss of generated concurrence is bounded by one half of the quantum Fisher information associated with the same interaction parameter.

Equivalently,
\begin{equation}
C(g,t)-\overline C(g,t)
\le
\frac{\sigma_g^2}{2}
F_Q^{(g)}(g,t)
\end{equation}
to leading order in the fluctuation variance. The same quantum Fisher information that determines the local distinguishability of states under changes in the interaction strength therefore also bounds the leading-order loss of entanglement generated by uncertainty in that interaction strength.

The saturation conditions derived in Ref.~\cite{SaleemCoE2025} immediately acquire an operational significance in the present context. Whenever the conditions for $\mathrm{CoE}(g,t)=F_Q^{(g)}(g,t)$ are satisfied, Eq.~(\ref{eq:RgCoE}) implies
\begin{equation}
\lim_{\sigma_g\rightarrow0}
\mathcal R_g(g,t)
=
\frac{1}{2}
F_Q^{(g)}(g,t).
\end{equation}
At such operating points, the quantum Fisher information completely determines the leading-order loss of entanglement caused by interaction-strength uncertainty.

This provides a practical way to use the result. In a two-qubit entangling experiment, one may estimate or bound \(F_Q^{(g)}\) for the intended interaction and combine it with an independently characterized fluctuation variance \(\sigma_g^2\). Equation~(\ref{eq:mainbound}) then bounds the expected reduction in concurrence without requiring a direct characterization of the concurrence over a range of fluctuating interaction strengths. At saturation, this estimate becomes exact to leading order in \(\sigma_g\). Thus, the saturation conditions identify operating points where information-geometric sensitivity is converted directly into entanglement-generation fragility, while the general inequality provides a robustness guarantee for arbitrary operating points in the weak-fluctuation regime.

More broadly, Eq.~(\ref{eq:mainbound}) demonstrates that information geometry constrains not only the precision with which an interaction parameter may be estimated, but also the robustness with which that interaction can generate entanglement.

\textit{Conclusion and outlook.}
We have investigated the robustness of entanglement generation in the presence of stochastic fluctuations of the interaction strength. For two interacting qubits, we introduced the robustness quantity \(\mathcal R_g\), which characterizes the reduction in concurrence per unit variance of the fluctuating interaction parameter. In the weak-fluctuation limit, we showed that \(\mathcal R_g\) reduces to one half of the curvature of entanglement, thereby providing a direct operational interpretation of the curvature introduced in Ref.~\cite{SaleemCoE2025}. Combining this relation with the previously established information-geometric bound on the curvature of entanglement yields the central result of this work: the fluctuation-normalized loss of generated concurrence is bounded by one half of the quantum Fisher information associated with the same interaction parameter.

Our results establish a direct connection between entanglement generation, imperfect control, and quantum metrology. While the quantum Fisher information is traditionally viewed as a measure of parameter-estimation sensitivity, we have shown that it also constrains the leading-order degradation of entanglement generated by uncertainty in that parameter. At operating points satisfying the saturation conditions of Ref.~\cite{SaleemCoE2025}, the quantum Fisher information completely determines the weak-fluctuation loss of entanglement, providing a direct route for estimating entanglement robustness from independently characterized noise and metrological sensitivity.

Several extensions of the present work merit investigation. An immediate direction is the generalization of the robustness bound to multipartite systems and many-body entangling dynamics, where the interplay between quantum Fisher information and entanglement generation is known to exhibit rich scaling behavior \cite{Pezze2018,Hyllus2012,Toth2012}. Another interesting direction is the extension to open quantum systems, where dissipation and decoherence may modify both the curvature of entanglement and the relevant notions of quantum Fisher information. Finally, it would be interesting to explore applications in quantum networking and distributed sensing protocols, where interaction-strength uncertainty, phase noise, and imperfect entanglement generation constitute fundamental performance limitations \cite{Kimble2008,Zang2024Distributed}.

\textit{Acknowledgement} Z.H.S will like to thank Stephen Gray and Anil Shaji for very insightful conversations on the topic. Z.H.S will also thank the grant Quantized 2.0 for supporting this work.

\bibliographystyle{apsrev4-2}
\bibliography{bibliography}

\end{document}